\documentclass[a4paper]{article}
    
    \usepackage{INTERSPEECH2021}
    \usepackage{multirow}
    
    \title{Serialized Multi-Layer Multi-Head Attention for Neural Speaker Embedding}
    \name{Hongning Zhu$^{1,2}$, Kong Aik Lee$^3$, Haizhou Li$^2$}
    \address{
      $^1$School of Computing, National University of Singapore, Singapore\\
      $^2$Department of Electrical and Computer Engineering, National University of Singapore, Singapore\\
      $^3$Institute for Infocomm Research, A$^{\star}$STAR, Singapore}
    \email{hongning\_zhu@u.nus.edu, lee\_kong\_aik@i2r.a-star.edu.sg, haizhou.li@nus.edu.sg}
    
\begin{document}
    
    \maketitle
    \begin{abstract}
    This paper proposes a serialized multi-layer multi-head attention for neural speaker embedding in text-independent speaker verification. In prior works, frame-level features from one layer are aggregated to form an utterance-level representation.  Inspired by the Transformer network, our proposed method utilizes the hierarchical architecture of stacked self-attention mechanisms to derive refined features that are more correlated with speakers. Serialized attention mechanism contains a stack of self-attention modules to create fixed-dimensional representations of speakers. Instead of utilizing multi-head attention in parallel, the proposed serialized multi-layer multi-head attention is designed to aggregate and propagate attentive statistics from one layer to the next in a serialized manner. In addition, we employ an input-aware query for each utterance with the statistics pooling.  With more layers stacked, the neural network can learn more discriminative speaker embeddings. Experiment results on VoxCeleb1 dataset and SITW dataset show that our proposed method outperforms other baseline methods, including x-vectors and other x-vectors + conventional attentive pooling approaches by 9.7\% in EER and 8.1\% in DCF$10^{-2}$.
     
    \end{abstract}
    
    \section{Introduction}
    
    Speaker verification aims to determine whether the speaker's identity of a test utterance is the same as the reference speech. In recent years, extracting speaker embeddings for speaker verification task has become the mainstream method. We have also witnessed a surge of interest in deep neural network (DNN) speaker embeddings \cite{lee2020two}. These DNN-based embeddings achieve superior performance than i-vector \cite{dehak2010ivector} and has become state-of-the-art methods. The key of speaker embeddings is to find fixed-dimensional representations that can represent the voice characteristics of speakers.
    
    Most deep neural networks for speaker embeddings consist of three components: (1) a DNN front-end for extracting frame-level features, (2) temporal aggregation of frame-level features to form utterance-level statistics, and (3) a speaker classifier followed by multi-class cross-entropy loss. The DNN front-end can be a time-delay neural network (TDNN) \cite{snyder2018x}, convolutional neural network (CNN) \cite{nagrani2017voxceleb, chung2018voxceleb2}, recurrent neural network (RNN)  \cite{heigold2016end, novoselov2018deep}, or other neural network architectures  \cite{snyder2016deep, safari2020self}. Pooling layers like global average pooling and statistics pooling are the most frequently used method for the temporal aggregation of frame-level features. It maps a variable-length sequence into a fixed-length representation. In x-vector embedding \cite{snyder2018x}, its statistics pooling comprises the mean and standard deviation of frame-level features. A plain-vanilla statistics pooling assigns equal weights to each frame-level feature, which ignores the importance of some critical frames. Numerous recent works \cite{okabe2018attentive, zhu2018self, india2019self} have proposed attention-based techniques to address this problem. However, there is a common limitation in these prior methods that they only use one simple pooling layer to aggregate frame-level features to form a fixed-length vector. 
    
    In \cite{vaswani2017attention}, it was shown that a stacked self-attention mechanism could achieve state-of-the-art results in natural language processing tasks \cite{devlin2018bert, dai2019transformer, strubell2018linguistically}. In addition to the stacked self-attention at the encoder, the Transformer network \cite{vaswani2017attention} uses a multi-head topology, where attention functions are performed in parallel at each attention layer. However, the stacked self-attention modules compose a serialized architecture of the Transformer network, where it can learn from previous layers with attentive information. With a deeper utterance-level aggregation network, the speaker embeddings will have greater capacity and become more discriminative. We conjecture that, if multi-head topology is applied in a serialized manner, more layers can be stacked and capture more robust speaker characteristics.

    In this paper, we present a method to extract speaker embedding, using a novel serialized multi-layer multi-head attention. We show that this method can capture representations that contain important information. With an input-aware query, the weighted statistics from different layers are calculated with self-attention, then fused into features of the next layer with temporal context. They are also aggregated in a serialized manner as the utterance-level speaker embedding for speaker verification. To the best of our knowledge, this study is the first to aggregate frame-level features into utterance-level embeddings with multi-layer attention network. We conduct experiments on VoxCeleb1 \cite{nagrani2020voxceleb} dataset and Speakers in the Wild (SITW) \cite{sitw} dataset. The experiments show the effectiveness over other conventional attentive pooling approaches.
    
    The remainder of the paper is organized as follows. Section 2 reviews two conventional attentive pooling approaches. Section 3 describes the proposed serialized attention framework. The dataset and experimental setup are presented in Section 4. We discuss the results of these experiments in Section 5. The conclusion is finally drawn in Section 6.

    \section{Attention in Neural Speaker Embedding}
    Neural speaker embeddings are fixed-dimensional representations of speech utterances extracted with DNNs, among which x-vector \cite{snyder2018x} is the most widely used. In x-vector, temporal aggregation is used to convert frame-level features into a single fixed-dimensional vector. Then a fully-connected layer maps the utterance-level features to a speaker embedding. It is believed that certain frames are more unique and important for discriminating speakers than others. Instead of assigning equal weights to each frame, an attention mechanism is often applied to obtain a set of weights as the importance of each frame over the input sequence.
    
    \subsection{Statistics pooling}
    Let $\mathbf{h}_t$ be the latent vector at the output of the frame processor network. By statistics pooling, we compute the mean and standard deviation of $\mathbf{h}_t$ along the temporal axis, $t = 1, 2, ..., T$. In particular, the first and second-order statistics are computed as follows:
    \begin{equation}\label{mean}
         \boldsymbol{\mu}= \frac{1}{T}\sum_{t=1}^{T}\mathbf{h_{t}}
    \end{equation}
    \begin{equation}\label{std}
         \boldsymbol{{\sigma}} = \frac{1}{T}\sqrt{\sum_{t=1}^{T}\mathbf{h_{t}} \odot \mathbf{h_{t}} - \boldsymbol{\mu} \odot \boldsymbol{\mu}}
    \end{equation}
    where equal weights of $\alpha _t = \frac{1}{T}$ are assigned to all frames. The operator $\odot$ represents element-wise multiplication. The mean and standard deviation are concatenated as a fixed-dimensional representation and mapped to the speaker embedding vector via an affine transformation, typically, implemented with a fully-connected (FC) layer.
    
    \subsection{Attentive statistics pooling}
    Attentive statistics pooling \cite{okabe2018attentive} method aims to capture the temporal information focusing on the importance of frames. An attention model works in conjunction with the original embedding neural network and calculates a scalar score $e_{t}$ for each frame, as follows:
    \begin{equation}
        e_{t} = \mathbf{v}^Tf(\mathbf{Wh}_{t} + \mathbf{b}) + k
    \end{equation}
    where $f(\cdot )$ is a non-linear activation function, such as tanh or ReLU. The scores are normalized over all frames with a softmax function as follows:
    \begin{equation}
        \alpha _{t} = \frac{{\rm exp}(e_{t})}{\sum_{\tau}^{T}{\rm exp}(e_{\tau})}
    \end{equation}
    The normalized scores are then used as the weights in the pooling layer to calculate a weighted mean $\boldsymbol{\tilde{\mu}}$ and a weighted standard deviation $\boldsymbol{\tilde{\sigma}}$:
    \begin{equation}\label{mu}
        \boldsymbol{\tilde{\mu}}= \sum_{t=1}^{T}\alpha _{t}\mathbf{h}_{t}
    \end{equation}
    \begin{equation}\label{sigma}
        \boldsymbol{\tilde{\sigma}} = \sqrt{\sum_{t=1}^{T}\alpha _{t}\mathbf{h}_{t} \odot \mathbf{h}_{t} - \boldsymbol{\tilde{\mu}} \odot \boldsymbol{\tilde{\mu}}}
    \end{equation}
    Compared to (\ref{mean}) and (\ref{std}), the weights $\alpha _{t}$ in (\ref{mu}) and (\ref{sigma}) are no longer uniformly distributed across the temporal axis. In this way, the utterance-level representation can focus on important frames that promotes a better speaker discrimination.
    
    \subsection{Self-attentive pooling}
    In \cite{liu2018exploring}, the authors used a more rigorous formulation based on a \{value, key, query\} tuple to construct the so-called self-attentive pooling mechanism.
    
    Let $(\mathbf{v}_{t}, \mathbf{k}_{t}, \mathbf{q})$ be the \{value, key, query\} tuple. Here, $\mathbf{v}_{t}$ is the value vector with $d_{v}$ dimensions, $\mathbf{q}$ is a time-invariant query with $d_q$ dimensions, and $\mathbf{k}_{t}$ is the key vector with $d_k$ dimensions. In \cite{liu2018exploring}, the $(\mathbf{v}_{t}, \mathbf{k}_{t})$ is derived from different layers of the frame processor network, while $\mathbf{q}$ is a trainable parameter. The query vector maps the key vector sequence     $[{{\mathbf{k}_{1}}, \mathbf{k}_{2}, ..., \mathbf{k}_{T}}]$ to the weights $[{{\alpha}_{1}, {\alpha}_{2}, ..., {\alpha}_{T}}]$ via scaled dot-production attention and softmax function:
    \begin{equation}\label{softmax}
    \alpha _{t} = {\rm softmax}\left( \frac{\mathbf{q}\cdot \mathbf{k}_{t}}{\sqrt{d_{k}}} \right)
    \end{equation}
    Note that the softmax is performed along the temporal axis. Finally, the weighted mean $\boldsymbol{\tilde{\mu}}$ and weighted standard deviation $\boldsymbol{\tilde{\sigma}}$ are computed in the same way as (\ref{mu}) and (\ref{sigma}) with the weights $\alpha _{t}$ applied on the value vectors $\mathbf{v}_{t}$.

    \section{Serialized Multi-head Attention}
    
\begin{figure}[htp]
    \centering %
    \includegraphics[width = 8cm]{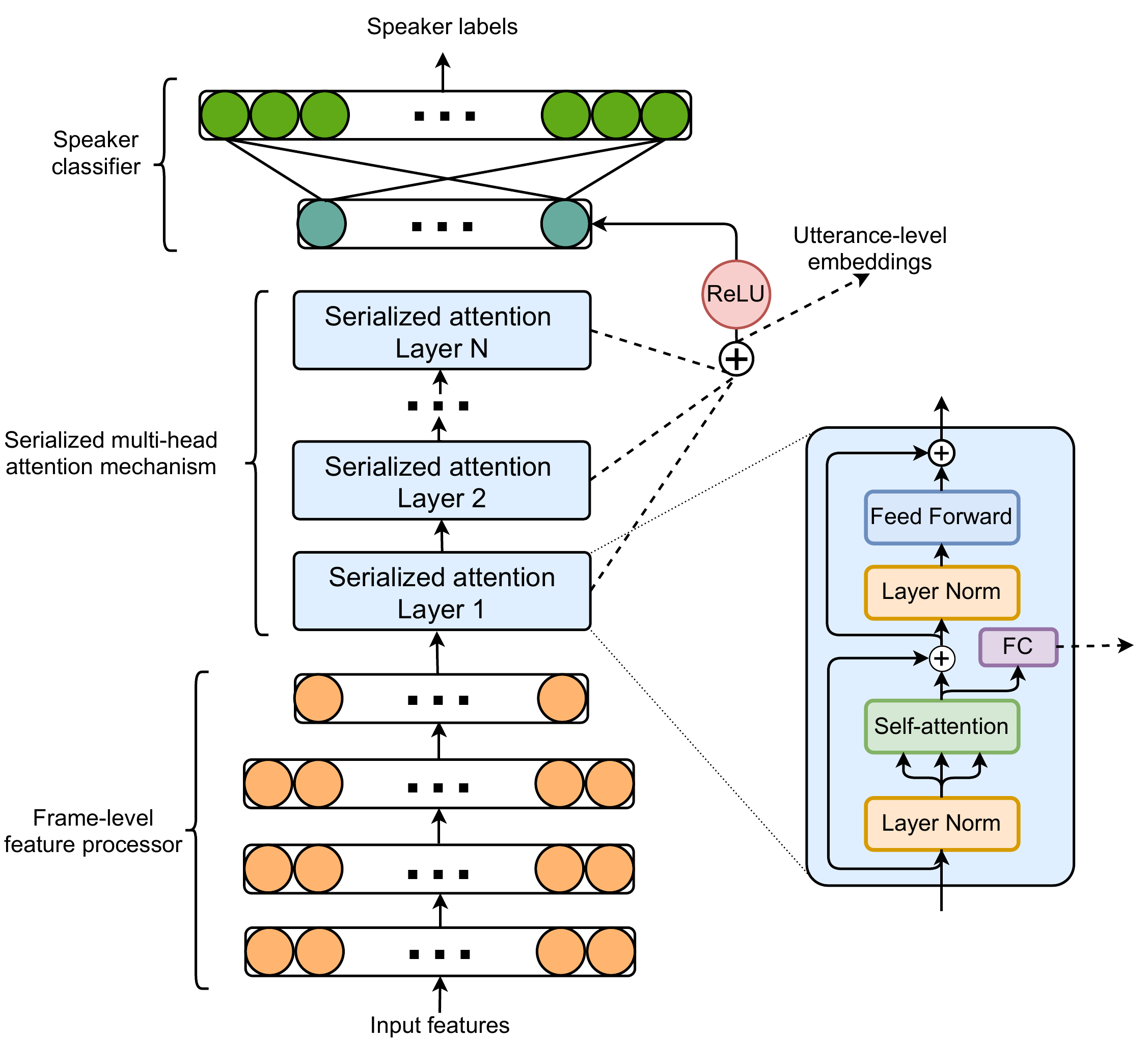}
    \caption{Deep speaker embedding neural network with a serialized multi-head attention mechanism.}
    \label{fig:architecture}
    \end{figure}

    In this section, we introduce the proposed serialized multi-layer multi-head attention mechanism. As depicted in Figure \ref{fig:architecture}, the embedding neural network consists of three main stages, namely, a frame-level feature processor, a serialized attention mechanism, and a speaker classifier. The frame-level feature processor is the same as that in the x-vector \cite{snyder2018x}, which uses a TDNN to extract high-level representations of the input acoustic features. The middle part of Figure \ref{fig:architecture}, a serialized attention mechanism is used to aggregate the variable-length feature sequence into a fixed-dimensional representation. The top part of Figure \ref{fig:architecture} are feed-forward classification layers. Similar to the x-vector, the entire network is trained to classify input sequences into speaker classes.
    
    \subsection{Serialized attention}
    The serialized attention mechanism consists of a stack of $N$ identical layers, and each layer is composed of two modules stacked together, i.e, a self-attention module and a feed forward module. We employ a residual connection around each of these modules. As in \cite{xiong2020layer}, layer normalization is applied on the input before the self-attention module and feed-forward module, separately. That is, the output of each sub-layer is $x + LayerNorm(Sublayer(x))$. 
    
    Instead of having multi-head attention in parallel, we propose to aggregate and propagate the information from one layer to the next in a serialized manner with stacked self-attention modules. In the original multi-head attention, the input sequence is split into several homogeneous sub-vectors called heads. However, a deeper architecture of the aggregation network will increase the representational capacity, with more discriminative features can be learned and aggregated at different levels. In the proposed serialized attention mechanism, self-attention module is performed in a serialized manner, allowing the model to aggregate information with temporal context from deeper layers. Skip connection is used throughout self-attention module for training a much deeper network \cite{oord2016wavenet} and aggregating serialized heads. Specifically, from the $n^{th}$ self-attention module ($n \in [1, ..., N]$), the weighted mean $\boldsymbol{\tilde{\mu }}$ and weighted standard deviation $\boldsymbol{\tilde{\sigma }}$ are obtained. After transformed by an affine transformation, it is converted to an utterance-level vector, also seen as a serialized head from layer $n$. The final utterance-level embedding is then obtained with the summation of the utterance-level vectors from all heads. After passing through a ReLU activation and Batch Normalization, it is then fed into classifier layers.
    
    \subsection{Input-aware self-attention}
\begin{figure}[htp]
    \centering %
    \includegraphics[width = 8cm]{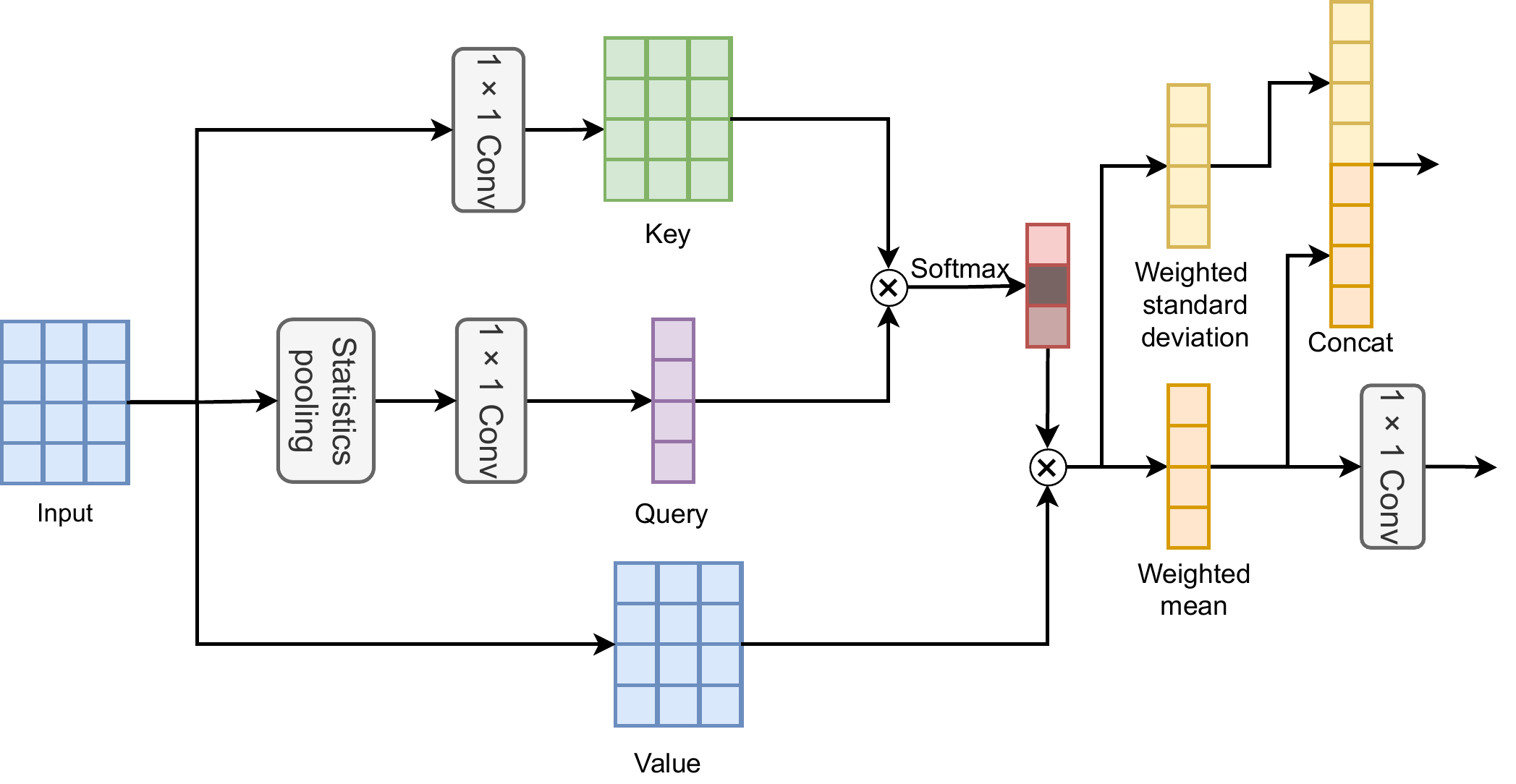}
    \caption{Self-attention mechanism with input-aware query.}
    \label{fig:attn}
    \end{figure}

    The attention function is mapping a query and a set of key-value pairs to an output \cite{vaswani2017attention}. Instead of using a fixed query for all utterances as in \cite{liu2018exploring}, we employ an input-aware unique query for each utterance. Considering that mean and standard deviation are capable of capturing the overall information and speech dynamics over an utterance, we use statistics pooling here to generate the query. As shown in Figure \ref{fig:attn}, consider an input sequence $[\mathbf{h}_{1}, \mathbf{h}_{2}, ..., \mathbf{h}_{T}]$ with $\mathbf{h}_{t} \in \mathbb{R}^{d}$, where $T$ is the length of the input sequence. The model transforms the input sequence into the query $\mathbf{q}$ as follows:
    \begin{equation}
      \mathbf{q} = \mathbf{W}_qg(\mathbf{h}_t)
    \end{equation}
    where $g(\cdot )$ is statistics pooling illustrated in Section 2.1, which is applied to calculate $[\boldsymbol{\mu}, \boldsymbol{\sigma}]$ with (\ref{mean}) and (\ref{std}), and $\mathbf{W}_q \in \mathbb{R}^{d_k\times 2d}$ is a trainable parameter.
    
    As for key-value pairs, in order to reduce the number of model parameters, the input sequence $[\mathbf{h}_{1}, \mathbf{h}_{2}, ..., \mathbf{h}_{T}]$ is directly assigned to the value sequence $[\mathbf{v}_{1}, \mathbf{v}_{2}, ..., \mathbf{v}_{T}]$ of $d$ dimensions without any extra computation. The key vector $\mathbf{k}_t$ is obtained by a linear projection with a trainable parameter $\mathbf{W}_{k} \in \mathbb{R}^{d_k\times d}$:
    \begin{equation}
       \mathbf{k}_t = \mathbf{W}_{k}\mathbf{h}_t
    \end{equation}
    With $(\mathbf{v}_{t}, \mathbf{k}_{t}, \mathbf{q})$ as \{value, key, query\} tuple, the weights are computed via scaled dot-product attention as in (\ref{softmax}). The first and second-order statistics are calculated the same as in (\ref{mu}) and (\ref{sigma}). The weighted mean vector $\boldsymbol{\tilde{\mu }}$ is then added to all frames after an affine transformation in the residual connection. 
    
    \subsection{Serialized multi-head embedding}
    After the self-attention module, the output from each self-attention layer is fed into a feed-forward module, which is to process the output to better fit the input for the next self-attention layer. The structure of the feed-forward module is the same as \cite{vaswani2017attention}. It consists of two linear transformations with a ReLU activation in between.
    \begin{equation}
        FFW(\mathbf{h}) =\mathbf{W}_{2}f(\mathbf{W}_{1}\mathbf{h} + \mathbf{b}_{1}) + \mathbf{b}_{2}
    \end{equation}
    where $\mathbf{h}$ is the input, $f(\cdot )$ is a ReLU function, and the linear transformations are different from layer to layer. $\mathbf{W}_{1} \in \mathbb{R}^{d_{ff}\times d}$, $\mathbf{W}_{2}\in \mathbb{R}^{d\times d_{ff}}$ with inner dimension $d_{ff}$, which can also be described as two convolutions with kernel size 1.
    
    The utterance-level embedding of serialized attention mechanism is fed into one fully-connected layer and a standard softmax layer. The softmax layer with each of the nodes corresponds to the speaker labels in the training set. Hence, the system is trained as a speaker classifier. For training, we employ back-propagation with cross entropy loss. Once the system is trained, the utterance-level embedding is used as the serialized multi-head embedding for the speaker verification task.

    \section{Experiment}

    \begin{table*}[]
    \caption{Performance on SITW-dev and SITW-eval. \textbf{Boldface} denotes the best performance for each column.}
    \setlength{\tabcolsep}{2.5mm}{
    \begin{tabular}{ccccccccc}
    \toprule
    \multirow{2}{*}{\textbf{Embedding}} & \multirow{2}{*}{\textbf{Dim.}} & \multirow{2}{*}{\textbf{\# params}} & \multicolumn{3}{c}{\textbf{SITW-dev}}                                  & \multicolumn{3}{c}{\textbf{SITW-eval}}                                 \\ \cline{4-9} 
                                        &                                &                                     & \textbf{EER(\%)} & \textbf{DCF$10^{-2}$} & \textbf{DCF$10^{-3}$} & \textbf{EER(\%)} & \textbf{DCF$10^{-2}$} & \textbf{DCF$10^{-3}$} \\ \hline
    statistic pooling \cite{snyder2018x}                  & 512                            & 4.47M                               & 2.81             & 0.294                    & 0.464                    & 3.25             & 0.324                    & 0.522                    \\
    attentive statistic \cite{okabe2018attentive}                 & 512                            & 4.48M                               & 2.54             & 0.276                    & 0.462                    & 3.17             & 0.304                    & 0.491                    \\
    self-attentive \cite{liu2018exploring}                     & 512                            & 4.73M                                & 2.77             & 0.288                    & 0.447                    & 3.09             & 0.305                    & \textbf{0.486}           \\ \hline
    \textbf{ours ($N = 4$)}                         & 256                            & 3.88M                                 & 2.54                 & 0.283                       & 0.478                         & 3.09                 & 0.314                         & 0.515                         \\
    \textbf{ours ($N = 5$)}             & 256       & 4.44M                      & 2.27         & 0.286         & 0.479         & 2.98          & 0.308      & 0.519                    \\
    \textbf{ours ($N = 6$)}                         & 256                            & 4.99M                               & \textbf{2.16}    & \textbf{0.265}           & \textbf{0.441}           & \textbf{2.82}    & \textbf{0.291}           & 0.499         \\
    \bottomrule
    \end{tabular}}
    \label{sitw}
    \end{table*}
    
    \begin{table*}[]
    \caption{Performance on VoxCeleb1-H and VoxCeleb1-E. \textbf{Boldface} denotes the best performance for each column.}
    \setlength{\tabcolsep}{2.5mm}{
    \begin{tabular}{ccccccccc}
    \toprule
    \multirow{2}{*}{\textbf{Embedding}} & \multirow{2}{*}{\textbf{Dim.}} & \multirow{2}{*}{\textbf{\# params}} & \multicolumn{3}{c}{\textbf{VoxCeleb1-H}}                         & \multicolumn{3}{c}{\textbf{VocCeleb1-E}}                         \\ \cline{4-9} 
                                        &                                &                                     & \textbf{EER(\%)} & \textbf{DCF$10^{-2}$} & \textbf{DCF$10^{-3}$} & \textbf{EER(\%)} & \textbf{DCF$10^{-2}$} & \textbf{DCF$10^{-3}$} \\ \hline
    statistic pooling \cite{snyder2018x}                  & 512                            & 4.47M                               & 4.50             & 0.425                   & 0.660                    & 2.57             & 0.283                    & 0.497                    \\
    attentive statistic \cite{okabe2018attentive}                & 512                            & 4.48M                               & 4.41             & 0.421                   & 0.667                    & 2.51             & 0.279                    & 0.493                    \\
    self-attentive \cite{liu2018exploring}                     & 512                            & 4.73M                                & 4.40             & 0.420                   & 0.666                    & 2.52             & 0.279                    & 0.465                    \\ \hline
    \textbf{ours ($N = 4$)}                         & 256                            & 3.88M                                    & 4.28                 & 0.407                         & 0.640                         & 2.48                 & 0.265                         & 0.467                         \\
    \textbf{ours ($N = 5$)}                         & 256                            & 4.44M                                    & 4.07                  & 0.380                         & 0.619                          & 2.38                 & 0.253                          & 0.452                         \\
    \textbf{ours ($N = 6$)}                         & 256                            & 4.99M                               & \textbf{3.99}    & \textbf{0.375}           & \textbf{0.616}           & \textbf{2.36}    & \textbf{0.242}           & \textbf{0.431}      \\
    \bottomrule
    \end{tabular}}
    \label{vox}
    \end{table*}

    \subsection{Dataset}
    The training set comprises the VoxCeleb2 \cite{chung2018voxceleb2} development dataset and it is an augmented version \cite{snyder2018x} with 5,994 speakers and 1,092,009 utterances.
    
    Our experiment consists of four test sets from two different datasets: (1) SITW development core-core condition \cite{sitw}; (2) SITW evaluation core-core condition; (3) the extended VoxCeleb1-E test set, which uses the entire Voxceleb1 \cite{nagrani2020voxceleb} (both train and test splits), containing 1,251 speakers and 579,818 random pairs; (4) the hard VoxCeleb1-H test set, which consists of 550,894 pairs with the same nationality and gender, sampled from the entire VoxCeleb1 dataset. The utterances in these test sets vary in length from 4 seconds to 180 seconds. Hence, we could evaluate all models with both short and long utterances.
    \subsection{Experiments setup}
    The acoustic features used in all experiments are a combination of 23-dimensional Kaldi MFCC feature with a frame-length of 25ms and 3-dimensional pitch. The features are applied cepstral mean-normalization (CMN) with a sliding window of 3 seconds. A voice activity detection (VAD) is then used to remove silence frames. 

    To compare the proposed approach with other speaker embedding systems, we pick three baselines based on x-vector proposed in \cite{snyder2018x}: (i) statistics pooling as in the traditional x-vector. (ii) attentive statistics pooling proposed in \cite{okabe2018attentive}. (iii) self-attentive pooling \cite{liu2018exploring}, where the output of 4-th layer with 1-layer transformation network of 500 nodes is used as the keys. For frame-level layers in these x-vector systems, there are 512 channels in each of the first four TDNN layers, while 1500 in the fifth. ReLU is used as the non-linear activation for all the systems.
    
    For the configuration of the proposed serialized attention mechanism, we use $d = 256$ and $d_k = 128$ in self-attention module, and the feed-forward dimension of $d_{ff} = 512$. To regularize the model, a dropout \cite{srivastava2014dropout} of 0.1 is applied on the output of each module, before it is added to the input. For frame-level feature presscessor, the first three TDNN layers are the same as the x-vector system. The forth layer projects the dimension of frame-level features from 512 to 256 without non-linearity and Batch Normalization for a smaller number of parameters. The dimension of the first speaker classification layer is 256, which is equal to the speaker embedding dimension. 
    
    We use the ASV-Subtools \cite{xmuspeech} and the Kaldi toolkits \cite{povey2011kaldi} to implement the speaker embedding networks. During training, each utterance is chunked to a vector sequence of 200 frames, and the batch size is set to 512. We use the AdamW optimizer \cite{loshchilov2017decoupled} with an initial learning rate of $10^{-3}$, and decrease the learning rate every epoch for 9 epochs. 
    After training the neural speaker embedding network, the Kaldi toolkit is used to train the PLDA model and the dataset used is the same as the training set. Before using PLDA, we use Linear Discriminant Analysis (LDA) to decrease the dimensionality to 128. Then the extracted embeddings are length-normalized and computed verification scores by the PLDA.
    
    \section{Results}
    In this section, we report the performance of serialized attention. We evaluate the experimental results in terms of Equal Error Rate (EER) and the minimum Decision Cost Function where prior target probability is set as 0.01 (DCF$10^{-2}$) and 0.001 (DCF$10^{-3}$).
    \subsection{Results on SITW}
    Table \ref{sitw} presents the performance on SITW. Although there is a trade-off between performance and network size, we can obtain the best performance with $N = 6$ with 1.11M more parameters compared to $N = 4$. In comparison with baseline methods, the serialized attention mechanism achieves better performance on both development set and evaluation set, which improves by nearly 14.96\% and 8.73\% in EER on the dev and eval sets, respectively. Although self-attentive pooling offers the best performance of DCF$10^{-3}$ on SITW-eval, our serialized attention achieves better performance in terms of EER and DCF$10^{-2}$. 
    
    \subsection{Results on VoxCeleb1}
    Table \ref{vox} shows the performance on two VoxCeleb1 test sets. Here, the proposed method outperforms all the baseline methods by a significant margin with fewer or comparable parameters. The serialized attention with 6 layers has shown the best results of all the evaluated approaches. Comparing to statistics pooling, the 6-layer serialized attention achieves 11.33\%, 11.76\%, and 6.67\%, relative improvements in terms of EER, DCF$10^{-2}$ and DCF$10^{-3}$, respectively, in VoxCeleb1-H set. In VoxCeleb1-E set, relative improvements of 8.17\%, 14.49\%, 13.28\% are also obtained in terms of EER, DCF$10^{-2}$ and DCF$10^{-3}$, respectively. Besides, the 4-layer serialized multi-head attention is able to perform better than other single-layer attention-based pooling methods with 13\% less parameters. It can be observed that stacking more layers will achieve further improvement. The results show that deeper serialized attention network makes the speaker embedding more discriminative. 
    
    
    \section{Conclusion}
    
    In this paper, we proposes a new method to extract speaker embedding with a serialized multi-layer multi-head attention mechanism. Serialized attention mechanism contains a stack of self-attention modules to create fixed-dimensional representations for speaker verification. Unlike previous attention-based methods, weighted statistics from different layers are aggregated in a serialized manner. With an input-aware query, the proposed mechanism can capture more discriminative features. By increasing the stacked layers, consistent improvement is further obtained. Evaluating the proposed method on VoxCeleb1 and SITW dataset, our method outperforms other baseline methods by 9.7\% in EER and 8.1\% in DCF$10^{-2}$.
    
    \section{Acknowledgement}
    This research work is supported by Programmatic Grant No. A1687b0033 from the Singapore Government’s Research, Innovation and Enterprise 2020 plan (Advanced Manufacturing and Engineering domain).

    \bibliographystyle{IEEEtran}
    
    \bibliography{mybib}

\begin{thebibliography}{10}
\providecommand{\url}[1]{#1}
\csname url@samestyle\endcsname
\providecommand{\newblock}{\relax}
\providecommand{\bibinfo}[2]{#2}
\providecommand{\BIBentrySTDinterwordspacing}{\spaceskip=0pt\relax}
\providecommand{\BIBentryALTinterwordstretchfactor}{4}
\providecommand{\BIBentryALTinterwordspacing}{\spaceskip=\fontdimen2\font plus
\BIBentryALTinterwordstretchfactor\fontdimen3\font minus
  \fontdimen4\font\relax}
\providecommand{\BIBforeignlanguage}[2]{{%
\expandafter\ifx\csname l@#1\endcsname\relax
\typeout{** WARNING: IEEEtran.bst: No hyphenation pattern has been}%
\typeout{** loaded for the language `#1'. Using the pattern for}%
\typeout{** the default language instead.}%
\else
\language=\csname l@#1\endcsname
\fi
#2}}
\providecommand{\BIBdecl}{\relax}
\BIBdecl

\bibitem{lee2020two}
K.~A. Lee, O.~Sadjadi, H.~Li, and D.~Reynolds, ``Two decades into speaker
  recognition evaluation-are we there yet?'' 2020.

\bibitem{dehak2010ivector}
N.~Dehak, P.~J. Kenny, R.~Dehak, P.~Dumouchel, and P.~Ouellet, ``Front-end
  factor analysis for speaker verification,'' \emph{IEEE Transactions on Audio,
  Speech, and Language Processing}, vol.~19, no.~4, pp. 788--798, 2010.

\bibitem{snyder2018x}
D.~Snyder, D.~Garcia-Romero, G.~Sell, D.~Povey, and S.~Khudanpur, ``X-vectors:
  Robust dnn embeddings for speaker recognition,'' in \emph{2018 IEEE
  International Conference on Acoustics, Speech and Signal Processing
  (ICASSP)}.\hskip 1em plus 0.5em minus 0.4em\relax IEEE, 2018, pp. 5329--5333.

\bibitem{nagrani2017voxceleb}
A.~Nagrani, J.~S. Chung, and A.~Zisserman, ``Voxceleb: a large-scale speaker
  identification dataset,'' \emph{arXiv preprint arXiv:1706.08612}, 2017.

\bibitem{chung2018voxceleb2}
J.~S. Chung, A.~Nagrani, and A.~Zisserman, ``Voxceleb2: Deep speaker
  recognition,'' \emph{arXiv preprint arXiv:1806.05622}, 2018.

\bibitem{heigold2016end}
G.~Heigold, I.~Moreno, S.~Bengio, and N.~Shazeer, ``End-to-end text-dependent
  speaker verification,'' in \emph{2016 IEEE International Conference on
  Acoustics, Speech and Signal Processing (ICASSP)}.\hskip 1em plus 0.5em minus
  0.4em\relax IEEE, 2016, pp. 5115--5119.

\bibitem{novoselov2018deep}
S.~Novoselov, A.~Shulipa, I.~Kremnev, A.~Kozlov, and V.~Shchemelinin, ``On deep
  speaker embeddings for text-independent speaker recognition,'' \emph{arXiv
  preprint arXiv:1804.10080}, 2018.

\bibitem{snyder2016deep}
D.~Snyder, P.~Ghahremani, D.~Povey, D.~Garcia-Romero, Y.~Carmiel, and
  S.~Khudanpur, ``Deep neural network-based speaker embeddings for end-to-end
  speaker verification,'' in \emph{2016 IEEE Spoken Language Technology
  Workshop (SLT)}.\hskip 1em plus 0.5em minus 0.4em\relax IEEE, 2016, pp.
  165--170.

\bibitem{safari2020self}
P.~Safari, M.~India, and J.~Hernando, ``Self-attention encoding and pooling for
  speaker recognition,'' \emph{arXiv preprint arXiv:2008.01077}, 2020.

\bibitem{okabe2018attentive}
K.~Okabe, T.~Koshinaka, and K.~Shinoda, ``Attentive statistics pooling for deep
  speaker embedding,'' \emph{arXiv preprint arXiv:1803.10963}, 2018.

\bibitem{zhu2018self}
Y.~Zhu, T.~Ko, D.~Snyder, B.~Mak, and D.~Povey, ``Self-attentive speaker
  embeddings for text-independent speaker verification.'' in
  \emph{Interspeech}, vol. 2018, 2018, pp. 3573--3577.

\bibitem{india2019self}
M.~India, P.~Safari, and J.~Hernando, ``Self multi-head attention for speaker
  recognition,'' \emph{arXiv preprint arXiv:1906.09890}, 2019.

\bibitem{vaswani2017attention}
A.~Vaswani, N.~Shazeer, N.~Parmar, J.~Uszkoreit, L.~Jones, A.~N. Gomez,
  L.~Kaiser, and I.~Polosukhin, ``Attention is all you need,'' \emph{arXiv
  preprint arXiv:1706.03762}, 2017.

\bibitem{devlin2018bert}
J.~Devlin, M.-W. Chang, K.~Lee, and K.~Toutanova, ``Bert: Pre-training of deep
  bidirectional transformers for language understanding,'' \emph{arXiv preprint
  arXiv:1810.04805}, 2018.

\bibitem{dai2019transformer}
Z.~Dai, Z.~Yang, Y.~Yang, J.~Carbonell, Q.~V. Le, and R.~Salakhutdinov,
  ``Transformer-xl: Attentive language models beyond a fixed-length context,''
  \emph{arXiv preprint arXiv:1901.02860}, 2019.

\bibitem{strubell2018linguistically}
E.~Strubell, P.~Verga, D.~Andor, D.~Weiss, and A.~McCallum,
  ``Linguistically-informed self-attention for semantic role labeling,''
  \emph{arXiv preprint arXiv:1804.08199}, 2018.

\bibitem{nagrani2020voxceleb}
A.~Nagrani, J.~S. Chung, W.~Xie, and A.~Zisserman, ``Voxceleb: Large-scale
  speaker verification in the wild,'' \emph{Computer Speech \& Language},
  vol.~60, p. 101027, 2020.

\bibitem{sitw}
M.~McLaren, L.~Ferrer, D.~Castan, and A.~Lawson, ``The speakers in the wild
  (sitw) speaker recognition database.'' in \emph{Interspeech}, 2016, pp.
  818--822.

\bibitem{liu2018exploring}
Y.~Liu, L.~He, W.~Liu, and J.~Liu, ``Exploring a unified attention-based
  pooling framework for speaker verification,'' in \emph{2018 11th
  International Symposium on Chinese Spoken Language Processing
  (ISCSLP)}.\hskip 1em plus 0.5em minus 0.4em\relax IEEE, 2018, pp. 200--204.

\bibitem{xiong2020layer}
R.~Xiong, Y.~Yang, D.~He, K.~Zheng, S.~Zheng, C.~Xing, H.~Zhang, Y.~Lan,
  L.~Wang, and T.~Liu, ``On layer normalization in the transformer
  architecture,'' in \emph{International Conference on Machine Learning}.\hskip
  1em plus 0.5em minus 0.4em\relax PMLR, 2020, pp. 10\,524--10\,533.

\bibitem{oord2016wavenet}
A.~v.~d. Oord, S.~Dieleman, H.~Zen, K.~Simonyan, O.~Vinyals, A.~Graves,
  N.~Kalchbrenner, A.~Senior, and K.~Kavukcuoglu, ``Wavenet: A generative model
  for raw audio,'' \emph{arXiv preprint arXiv:1609.03499}, 2016.

\bibitem{srivastava2014dropout}
N.~Srivastava, G.~Hinton, A.~Krizhevsky, I.~Sutskever, and R.~Salakhutdinov,
  ``Dropout: a simple way to prevent neural networks from overfitting,''
  \emph{The journal of machine learning research}, vol.~15, no.~1, pp.
  1929--1958, 2014.

\bibitem{xmuspeech}
F.~Tong, M.~Zhao, J.~Zhou, H.~Lu, Z.~Li, L.~Li, and Q.~Hong, ``Asv-subtools:
  Open source toolkit for automatic speaker verification,'' in \emph{ICASSP
  2021 - 2021 IEEE International Conference on Acoustics, Speech and Signal
  Processing (ICASSP)}, 2021, pp. 6184--6188.

\bibitem{povey2011kaldi}
D.~Povey, A.~Ghoshal, G.~Boulianne, L.~Burget, O.~Glembek, N.~Goel,
  M.~Hannemann, P.~Motlicek, Y.~Qian, P.~Schwarz \emph{et~al.}, ``The kaldi
  speech recognition toolkit,'' in \emph{IEEE 2011 workshop on automatic speech
  recognition and understanding}, no. CONF.\hskip 1em plus 0.5em minus
  0.4em\relax IEEE Signal Processing Society, 2011.

\bibitem{loshchilov2017decoupled}
I.~Loshchilov and F.~Hutter, ``Decoupled weight decay regularization,''
  \emph{arXiv preprint arXiv:1711.05101}, 2017.

\end{thebibliography}
    
    
    \end{document}